\documentclass[preprint,12pt]{aastex}

\newcommand{\NGC}[1]{\objectname[NGC]{NGC~{#1}}}

\shorttitle{LLAGNs at High Resolution}
\shortauthors{Anderson, Ulvestad, \& Ho}

\begin{document}


\title{Low-Luminosity AGNs at the Highest Resolution: Jets or
  Accretion Flows?}


\author{James M. Anderson\altaffilmark{1,2}}
\email{janderso@nrao.edu}
\author{James S. Ulvestad\altaffilmark{1}}
\email{julvesta@nrao.edu}
\and
\author{Luis C. Ho\altaffilmark{3}}
\email{lho@ociw.edu}






\altaffiltext{1}{National Radio Astronomy Observatory, P.O. Box O,
  1003 Lopezville Road, Socorro, NM 87801.}
\altaffiltext{2}{Department of Physics, New Mexico Institute of Mining
  and Technology, Socorro, NM 87801.}
\altaffiltext{3}{The Observatories of the Carnegie Institution of Washington, 813 Santa Barbara Street, Pasadena, CA 91101.}


\begin{abstract}
  Six low-luminosity active galactic nuclei have been imaged at
  multiple frequencies from 1.7--43~GHz (2.3--15~GHz for three of the
  galaxies) using the Very Long Baseline Array.  In spite of dynamic
  ranges of about 100 in several frequency bands, all six galaxies
  remain unresolved, with size limits at 8.4~GHz of $10^3$--$10^4$
  times the Schwarzschild radii of the black holes inferred at their
  galactic centers.  The galaxy spectra are roughly flat from 1.7 to
  43~GHz, rather than steepening to classical optically thin
  synchrotron spectra at high frequencies.  Although the spectral
  slopes somewhat resemble predictions for advection-dominated
  accretion flows, the luminosities are too high for the black hole
  masses of the galaxies and the slight spectral steepening at high
  frequencies cannot be explained by standard simple models of such
  accretion flows.  In contrast, compact jets can accommodate the
  average spectral index, the relatively high radio luminosity, and
  the unresolved appearance, but only if the jets in all six galaxies
  are fairly close to our line of sight.  This constraint is in
  agreement with inclination angle predictions for five of the six
  AGNs based on the dusty torus unification model.
\end{abstract}




\defcitealias{Ulvestad_H2001.0}{UH01a}
\defcitealias{Ulvestad_H2001.1}{UH01b}
\defcitealias{Ho_U2001}{HU01}
\defcitealias{Falcke_B1999}{FB99}
\defcitealias{Mahadevan_1997}{M97}

\section{Introduction\label{sec:intro}}


The physical processes that control the accretion and outflow in
low-luminosity active galactic nuclei (LLAGNs) are still not well
understood.  In their study of the Palomar Seyfert sample,
\citet*{Ho_FS1997.0} found that almost 40\% of bright ($B_T <
12.5$~mag) nearby galaxies have optical spectra indicating the
presence of an AGN.  Although their line emission resembles scaled
down versions of more luminous Seyfert nuclei, which are thought to be
driven by accretion in thin disks, LLAGN properties at other
wavelengths suggest that different processes are at work.  Whereas
``classical'' Seyfert galaxies have radio spectral indices $\alpha
\approx -0.7$, where $S_\nu \propto \nu^{+\alpha}$, almost half of the
low-luminosity Seyfert galaxies in the Palomar Seyfert sample have
flat or positive spectral indices \citetext{\citealp{Ulvestad_W1989};
  \citealp{Ho_U2001}, hereafter \citetalias{Ho_U2001};
  \citealp{Ulvestad_H2001.0}, hereafter
  \citetalias{Ulvestad_H2001.0}}.  Similar results have been found for
low-ionization nuclear emission regions (LINERs) and other AGNs in the
Palomar sample (\citealp{Nagar_ea2000,Nagar_ea2002};
\citealp*{Filho_BH2000,Filho_BH2002}).  These flat-spectrum objects
typically appear compact on arcsecond scales, and high resolution
studies with the Very Long Baseline Array\footnote{The VLBA is
  operated by the National Radio Astronomy Observatory, a facility of
  the National Science Foundation, operated under cooperative
  agreement by Associated Universities, Inc.}
\citep[VLBA;][]{Napier_ea1993} show that they remain compact even on
milliarcsecond scales \citep[hereafter
\citetalias{Ulvestad_H2001.1}]{Falcke_ea2000,Ulvestad_H2001.1}.

Following the work of \citet{Ferrarese_M2000},
\citet{Gebhardt_ea2000}, and others, it has become clear that both
normal and active galaxies with bulges have supermassive black holes
at their nuclear centers.  Accretion onto these black holes is thought
to fuel the AGN process for both ``normal'' and low-luminosity AGNs.
Since their black hole masses cover roughly the same range
\citep[see, for example, ][]{Ho_2002}, the mass of the central object
cannot account for the difference in emission levels.  The explanation
for why LLAGNs emit so little radiation may be either that they have
far less mass accreting onto them, or that the dominant mode of the
accretion mechanism itself must be different, or both.

The center of our own Milky Way Galaxy has been used as the testbed
for many models that attempt to explain the physics of LLAGNs.
Acceleration measurements of stellar orbits near the source
Sagittarius~A$^\ast$ (Sgr~A$^\ast$) leave little doubt as to the
existence of a supermassive black hole of $\sim 3.3\times
10^6~\mathrm{M_\Sun}$ at the Galactic Center
\citep{Ghez_ea2003,Schoedel_ea2003}.  Invisible in the optical and
near-infrared, Sgr~A$^\ast$ has a mildly inverted spectrum from 1 to
100~GHz \citep{Falcke_ea1998}.  The luminosity of Sgr~A$^\ast$ is
extremely low, even compared with other LLAGNs.  Advection dominated
accretion flows (ADAFs) have been used to explain the spectral energy
distributions (SEDs) of Sgr~A$^\ast$ and other LLAGNs during the past
decade (\citealp{Narayan_Y1994,Narayan_Y1995.1};
\citealp*{Narayan_YM1995,Narayan_MQ1998}).  In the ADAF model, a hot
plasma develops in which the protons decouple from the electrons,
allowing the protons to carry most of the energy released by the
accretion process into the black hole, limiting the amount that is
radiated away.  Hot thermal electrons generate synchrotron radiation
that is self-absorbed, resulting in a radio spectral index of $\sim
0.4$ \citep{Mahadevan_1997}.  Bremsstrahlung and inverse-Compton
processes generate additional emission, producing a relatively
well-defined spectrum from radio through X-ray frequencies.

An alternative hypothesis for the emission from LLAGNs is the
compact-jet model
\citep{Falcke_B1995,Falcke1996,Falcke_B1999,Falcke_M2000}.  Plasma is
ejected in a collimated beam away from the black hole, and expands
sideways to form a cone-shaped emission region.  Synchrotron emission
is initially self-absorbed, becoming progressively more transparent to
lower frequencies as the material travels away from the black hole,
producing an overall radio spectrum that can be quite flat.


With both ADAF (and other related low-radiation accretion flows) and
jet models failing to account for the overall SEDs in
LLAGNs (see, for example,
\citealp{Narayan_MQ1998}; \citealp*{Yuan_MF2002}), several attempts
have been made recently to combine the two models.
\citet{Yuan_MF2002} have reasonable success in explaining the spectrum
of Sgr~A$^\ast$ with this combined model.  Although this work has been
extended to \NGC{4258} \citep{Yuan_ea2002}, rigorous testing of a
large sample of LLAGNs in the nearby universe has yet to be performed.
These LLAGNs are many orders of magnitude \emph{more} luminous than
Sgr~A$^\ast$, but are also several orders of magnitude \emph{less}
luminous than ``classical'' AGNs.  They therefore occupy a luminosity
range that has thus far had few detailed studies.

We have formed an LLAGN sample from the \citetalias{Ho_U2001} study,
selecting objects which have 5~GHz Very Large Array (VLA) peak flux
densities between 5 and 30~mJy and spectral indices at milliarcsecond
resolution of $\alpha_{1.4}^{5} > -0.35$.  In this paper we discuss
simultaneous, multifrequency radio imaging of six LLAGNs.  The purpose
of these observations, and the point of the present paper, is to
measure the sizes and radio spectra of LLAGNs within
$10^4$~Schwarzschild radii of their black holes, and to determine what
constraints these parameters can place on possible ADAF and jet
models.

\section{Observations and Analysis\label{sec:obs}}

\NGC~{3147}, \NGC~{4203}, and \NGC{4579} were previously observed with
the VLBA at 1.7--8.4~GHz \citepalias{Ulvestad_H2001.1}.  Here we
present new VLBA observations of these galaxies at 8.4--43~GHz and
2.3--15~GHz observations of \NGC{4168}, \NGC{4235}, and \NGC{4450},
which have no previous VLBA observations.  Details of the
observations are presented in Table~\ref{tab:obs}.  Each galaxy was
observed in a single 10~hour run at a data recording speed of
256~$\mathrm{Mb~s^{-1}}$, with individual frequencies observed
multiple times in short time blocks spread throughout the full
10~hours to improve $(u,v)$ coverage.  The interleaving of observing
frequencies also results in a mean observing epoch which is nearly
simultaneous for all frequencies; this is important in order to
properly measure the spectral index, as LLAGNs are known to have
variable radio emission \citep[e.g.,][]{Nagar_ea2002}.  We applied an
amplitude calibration using a priori gain values together with system
temperatures measured during the observations; typically, this
calibration is accurate to within 5\% at the lower frequencies and
10\% for 22 and 43~GHz.

Initial clock and atmospheric (phase) errors were derived from the
calibrator sources listed in Table~\ref{tab:obs} using the technique
of phase-referencing \citep{Beasley_C1995}\footnote{The phase
  referencing involves repetitive switching between a strong
  calibrator source and the target galaxy in order to calibrate the
  atmospheric phase for the galaxy.  Our switching cycle ranged from
  4~minutes at the lower frequencies to 45~seconds at the higher
  frequencies.  Another bright, nearby check source is occasionally
  observed to test the effectiveness of the phase calibration.}.  This
initial calibration was used to determine the galaxy core positions
shown in Table~\ref{tab:gal_att}.  Uncertainties in the positions
generally are dominated by the uncertainties in the phase calibrator
positions, but contributions from ionospheric and tropospheric phase
fluctuations and residual phase errors can be important for some objects.

Each galaxy was detected at all frequencies in this initial imaging
process.  Phase-only self-calibration was then applied iteratively.
The resulting rms noise levels in the images far away from the galaxy
cores are consistent with predictions and vary from
40--70~$\mathrm{\mu Jy~beam^{-1}}$ for 2.3--8.4~GHz, up to about
500~$\mathrm{\mu Jy~beam^{-1}}$ at 43~GHz.  Beam widths range from
$\sim 6$~mas at 2.3~GHz, to $\sim 1.6$~mas at 8.4~GHz, to $\sim
0.4$~mas at 43~GHz using natural weighting.

Similar processing steps were performed on the substantially brighter
check sources, showing that all of the significant phase variations
were corrected in the target self-calibration procedures, except the
43~GHz observations of \NGC{3147} and \NGC{4203}, for which
self-calibration failed.  Therefore, the 43~GHz flux densities for
\NGC{3147} and \NGC{4203} have been scaled from the initial
phase-referenced values by the fractional increase in coherence seen
in the check sources and assigned a large uncertainty for potential
differences between the target and check source decoherence.

The integrated flux densities for the target galaxies as estimated by
Gaussian fitting are shown in Table~\ref{tab:flux}.  Since we have
only one overlapping frequency between \citetalias{Ulvestad_H2001.1}
and our higher frequency observations, we correct for possible
variability by shifting the low-frequency data for \NGC{3147},
\NGC{4203}, and \NGC{4579}, so that the 8.4~GHz flux densities are
identical at the two epochs and the spectra between $.14$ and
$8.4$~GHz are unchanged.  This amounts to an overall shift in
luminosity, with a maximum change of 19\% for \NGC{4579}.  Estimated
uncertainties from self-calibration and measurement errors have been
added in quadrature to the overall uncertainties in the amplitude
scale, in order to derive the final flux-density uncertainties.  The
LLAGN cores for these objects contain virtually all of the flux seen
on larger angular size scales with the VLA
(\citetalias{Ulvestad_H2001.1}; J.~M.  Anderson, et al., in
preparation), indicating that there is no significant emission from
large-scale features such as parsec-scale jets.

\section{Results\label{sec:results}}

\subsection{LLAGN Radio Images}

Our deep observations have resulted in images with signal-to-noise
levels of $\sim 100$ for frequencies of 2.3--15~GHz.  No significant
features reminiscent of jets were seen well above the noise level in
any of the six LLAGNs in this study.  \NGC{4579}, the brightest object
with the correspondingly highest signal-to-noise levels, has an
extension to the northeast visible at both 8.4 and 15~GHz, but this is
only seen at the 2-$\sigma$ level and represents less than 1\% of the
total flux in the image.  Figure~\ref{fig:8GHz_images} presents
8.4~GHz images of our LLAGNs; images at other wavelengths are similarly
pointlike.

Gaussian component fitting with the AIPS task JMFIT confirms the
visual impression that all of the VLBA images are dominated by a
single unresolved component.  Table~\ref{tab:gal_size_limits} shows
3-$\sigma$ upper limits to the major axes of the component sizes at
8.4~GHz in addition to the beam sizes.  The VLBA size upper limits
restrict the physics of the emission regions for these LLAGNs.
Table~\ref{tab:gal_size_limits} shows the resulting lower limits to
the brightness temperatures and upper limits to the \emph{radius} of
the emission region (in Schwarzschild radii, $R_\mathrm{S}$).

\subsection{LLAGN Core Shift}

The process of phase referencing not only allows weak sources to be
observed with VLBI techniques, it also allows the measurement of
accurate positions relative to the phase reference source.  This
process can, in theory, allow the detection of a core position shift
as a function of frequency.  Such a core shift is one of the
predictions of the compact-jet models \citep{Falcke_B1999}.  Our
measurements have sufficient signal-to-noise to permit measurement of
the core positions to $\ll 0.1$~mas.  Unfortunately, any core shift
within the phase calibrator, and, more importantly, phase fluctuations
due to the ionosphere and the troposphere can affect measurements to
detect core shift.  The measured core shifts prior to self-calibration
for our target and check sources are 0.2--0.5~mas between 8.4 and
15~GHz, extending to several milliarcseconds between 2.3 and 15~GHz
along roughly straight directions as the frequency is decreased.  The
angular size of the shift is roughly proportional to the square of the
wavelength, suggesting that the measured core shifts are due to
residual phase delays in the ionosphere (see
Figure~\ref{fig:core_shift}).  Models of the electron content of the
ionosphere for each observation were obtained from
\url{cddisa.gsfc.nasa.gov} \citep{Walker_C1999}; these models predict
ionosphere-induced position shifts between the phase calibrators and
other sources ranging from 0.2 to 0.5~mas between 8.4 and 15~GHz,
roughly in agreement with our measurements.  The directions of the
core shifts for the target and check source pairs are also consistent
with residual ionospheric delays.  We are therefore left with an upper
limit to any intrinsic core shift of $\sim 0.3$~mas between 8.4 and 15~GHz.

\subsection{LLAGN Radio Spectra}

Figure~\ref{fig:power_spectrum} shows the radio spectra for the six
LLAGNs in this paper.  The spectra are extremely flat.  In general,
the spectra rise slightly up to about 5--10~GHz then become flat or
even turn over slightly.  The 2.3--5.0~GHz spectral indices
($\alpha_{2.3}^{5.0}$) are remarkably uniform (see
Table~\ref{tab:gal_spec}); the 8.4--15~GHz spectral indices
($\alpha_{8.4}^{15}$) also are similar to one another, although there
are two outliers.  The high frequency data at 22 and 43~GHz, although
a bit noisy, indicate that the SEDs continue to be flat at high
frequencies.

The similarity among the spectra suggests that the physical mechanism
responsible for the radio emission is the same in all six objects.
The radio powers are all within a factor of 200 of one another, and
the inferred black hole masses are also within a factor of 20 of one
another (see Table~\ref{tab:gal_att}), with the more massive objects
having higher radio powers, so the general physical conditions in
these six objects would appear to be very similar.  We have combined
the individual spectra to make the mean SEDs shown in
Figure~\ref{fig:sed}.  After normalizing each galaxy spectrum to unity
at either 2.3 or 8.4~GHz, an unweighted mean was calculated for each
frequency, as well as the rms scatter about the mean.  The statistical
uncertainties in the mean values are about 10\% for all points in the
curve normalized at 8.4~GHz.

The galaxy spectra are all very similar at low frequencies, as
indicated by the small error bars at 1.7 and 5.0~GHz in the curve
normalized at 2.3~GHz in Figure~\ref{fig:sed}.  In contrast, the
behavior around 5 to 8~GHz is more varied, with some spectra
continuing to rise while others actually fall.  Then, around 15~GHz
the spectra become more similar to one another again as most of the
spectra become relatively flat.  The data suggest a phenomenon where
the spectra turn from slightly rising to being flat, with the turnover
frequency varying from one galaxy to the next.

The low-frequency portion of the mean spectrum rises modestly with
$\alpha_{1.7}^{8.4} = 0.16 \pm 0.08$ in the data normalized at
8.4~GHz.  The spectrum then turns over slightly (but is also
consistent with being completely flat) with $\alpha_{8.4}^{22} = -0.11
\pm 0.13$.  The overall spectrum is quite smooth --- a least squares
fit for a constant spectral index from 1.7 to 43~GHz gives
$\overline{\alpha} = 0.084 \pm 0.029$ with an rms scatter of only 7\%.

\section{Comparison with the ADAF Model\label{sec:ADAFs}}

Although ADAF models have had some success in explaining the infrared
through X-ray behavior of black hole systems, the standard ADAF models
have traditionally been less than satisfactory at explaining the
relatively strong radio emission as compared with the submillimeter
and X-ray emission (e.g.,
\citealp*{Manmoto_MK1997}; \citealp{Narayan_ea1998,Narayan_MQ1998}).
ADAF models have become less popular over the past few years, mostly
because they are unstable to convection and mass loss in winds, as
pointed out in the early paper by \citet{Narayan_Y1994}.
Modifications to ADAFs, such as the introduction of a nonthermal
electron distribution \citep*[e.g.,][]{Yuan_QN2003} to increase the
radio luminosity or the introduction of outflow winds
\citep[e.g.,][]{Blandford_B1999} to lower the effective accretion
rates, can improve fits to individual objects.  However, these models
are inevitably more complicated, and we defer comparisons with such
models to future publications.  Instead, we will concentrate here on
the simple ADAF model, which does illustrate the basic features of
radiatively inefficient accretion systems, and may play an important
role in some portions of the emission spectrum.

In the basic ADAF model
\citep[e.g.,][]{Narayan_Y1995.1,Narayan_MQ1998}, the accretion disk
surrounding the black hole is geometrically thick, with the height of
the disk on the order of the radius from the black hole.  Synchrotron
emission, radiated from thermal electrons in the accretion disk, is
self-absorbed at high frequencies close to the black hole, becoming
optically thin to progressively lower frequencies further away from
the black hole.  For the radio frequencies of interest here, the
typical size of the emission region is a few hundred Schwarzschild
radii.  This is considerably less than the upper limits to the sizes of
our LLAGNs shown in Table~\ref{tab:gal_size_limits}.  The distribution
of the emission on the sky does depend on the viewing angle somewhat,
but should be quite symmetric about the center of the black hole for
our observing frequencies. 
Therefore, the ADAF model predicts extremely small core shifts with
frequency, well within our upper limits.

A proper treatment of the emission spectrum from ADAFs requires a
detailed numerical computation.  We will present the results of such
calculations in a future paper, and for this paper we instead use the
scaling laws provided by \citet[hereafter
\citetalias{Mahadevan_1997}]{Mahadevan_1997}.  Although various
approximations used in the derivation of these scaling laws reduce
their accuracy, the luminosity results at high radio frequencies are
probably accurate to within a factor of 2 or so for the basic ADAF
model.  A power per unit frequency interval is given by
\citetalias{Mahadevan_1997}'s equation~25 in terms of the mass of the
black hole, the observing frequency, and the mass accretion rate
scaled to units of the Eddington accretion rate ($\dot{m}$).  Other
``variables'' explicit or implicit in this equation can be rewritten
in terms of these three parameters \citepalias[\S5 and the
Appendices of][]{Mahadevan_1997}.  Figure~\ref{fig:ADAF_43} shows the
predicted 43~GHz radio power as a function of accretion rate for the
black hole masses of our six LLAGNs.  Except for an overall scaling
factor, the curves are nearly all the same, as the ADAF radio power
dependence on black hole mass and accretion rate is nearly separable.
This figure illustrates the important fact that the basic ADAF model
predicts a maximum radio power at an accretion rate of about $\dot{m}
= 10^{-2.5}$.

This maximum power is less than the observed radio power, as shown in
Figure~\ref{fig:ADAF_spec}, where the maximum possible radio power in
the \citetalias{Mahadevan_1997} model is shown as a function of
frequency together with the observed spectra for \NGC{3147},
\NGC{4203}, and \NGC{4579} (results for the other three galaxies are
qualitatively similar).  The ADAF model spectra are generally more
than an order of magnitude below the observed spectra.  The only
exception is \NGC{3147}, where the model comes within a factor of 2.5
of the data at the highest frequencies.  The ADAF spectra have a
continuous spectral index of $\alpha = +0.4$, significantly larger
than the overall spectral indices of the sample
galaxies.\footnote{\citeauthor{Mahadevan_1997} notes that his model
  prediction for the radio power spectrum does not fully account for
  the ADAF dependence on radius, and suggests that the actual
  dependence should be closer to $P_{\!\nu} \propto \nu^{1/3}$, which
  could raise the radio luminosity somewhat.  The
  \citetalias{Mahadevan_1997} model also assumes that the ADAF is
  viewed pole-on.  We plan to produce numerical simulations of the
  ADAF model which account for such dependencies in a future paper.}
As shown by Table~\ref{tab:gal_spec}, most of the spectral indices at
low frequencies are reasonably consistent with the ADAF prediction,
but the observed spectra generally turn over slightly at higher
frequencies, despite the ADAF model predicting a continuous rise to
submillimeter frequencies.  Therefore, basic ADAFs probably make no
significant contribution to the radio output of our LLAGNs, except
possibly in \NGC{3147} where the slight rise at high frequencies may
or may not be due to an ADAF.  Modifications to include a population
of nonthermal electrons appear necessary to generate the observed
radio powers and spectral indices from ADAF-based systems.  We believe
that such modifications should allow ADAFs to remain a viable model
for LLAGN emission.

\section{Comparison with the Jet Model\label{sec:jets}}

\subsection{Model Comparison}


\citetalias{Ulvestad_H2001.1} concluded that the most natural way to
account for the SEDs of \NGC{3147}, \NGC{4203}, and \NGC{4579} in the
radio and X-ray was to invoke a jet to explain the radio emission in
combination with an ADAF to explain the X-ray emission.  Compact jet
models can easily accommodate the flux densities and spectral indices
of our LLAGNs.  Inspection of one jet model has interesting
implications for the physics of LLAGN accretion regions.

\citet[hereafter \citetalias{Falcke_B1999}]{Falcke_B1999} present a
set of simplified analytic equations for their jet model of radio
emission from black hole accretion systems based on the conical jet
model of \citet{Blandford_K1979}.  The model requires information
about the inclination of the jet axis to the line of sight ($i$, with
$i = 0\degr$ having the jet directed toward the observer) and the
characteristic angular size of the emission region
($\Phi_\mathrm{jet}$) in order to calculate the jet power
($Q_\mathrm{jet}$) and characteristic electron Lorentz factor
($\gamma_\mathrm{e}$).  Their model gives results consistent with
observations for LLAGN systems such as \objectname[]{Sgr~A$^\ast$} and
\NGC{4258}.  They suggest that a typical value of $\gamma_\mathrm{e}$
for LLAGNs may be $\sim 300$, and their equation~20 provides a means
to calculate the jet power in terms of the radio flux and inclination
angle.  By additionally using \citetalias{Falcke_B1999} equations 8
and~16, $\Phi_\mathrm{jet}$ can be calculated as a function of~$i$.


Figure~\ref{fig:jet_model}\textit{a} shows the jet model results for
the angular size of \NGC{4579} using three different values of
$\gamma_\mathrm{e}$.  The first model uses \citetalias{Falcke_B1999}
equation~20 with $\gamma_\mathrm{e} = 300$ (solid lines).  For small
inclination angles (jet pointed at the observer) the angular size is
small, while for large $i$ the angular size becomes very large.  The
expected inclination angle for a randomly oriented jet is $60\degr$,
which corresponds to $\Phi_\mathrm{jet} = 2.0$~mas at 8.4~GHz for
\NGC{4579}.  This is much larger than our upper limit from the VLBA
imaging of \NGC{4579}.  In order for this jet model to be valid, the
$0.54$~mas upper limit to the size requires $i \leq 40\degr$.  The
second model uses the parameters \citetalias{Falcke_B1999} derived for
\NGC{4258}, where a jet \emph{is} seen by the VLBA
\citep[e.g.][]{Herrnstein_ea1998}, with $\gamma_\mathrm{e} = 630$
(dotted lines).  For most inclination angles, $\Phi_\mathrm{jet}$ is
much larger in the $\gamma_\mathrm{e} = 630$ model compared to the
$\gamma_\mathrm{e} = 300$ model, and the $0.54$~mas upper limit in
size further restricts the inclination to $i \leq 27\degr$.  Although
decreasing $\gamma_\mathrm{e}$ from the moderate value of 300 does
slightly decrease $\Phi_\mathrm{jet}$ at small inclination angles, a
singular point in the \citetalias{Falcke_B1999} equations (resulting
from simplifications \citetalias{Falcke_B1999} made to their formulae)
produces extremely large $Q_\mathrm{jet}$ values, and hence
$\Phi_\mathrm{jet}$ values, at progressively smaller inclination
angles as $\gamma_\mathrm{e}$ is reduced.  A small decrease to
$\gamma_\mathrm{e} = 250$ (dashed lines) causes the model to predict
infinite size at $i=56\degr$.  Thus, the $\gamma_\mathrm{e} = 300$
model closely represents the least restrictive model on the allowable
inclination angle range.

In the \citetalias{Falcke_B1999} model, ``the characteristic size
scale of the core region is actually equivalent to the offset of the
radio core center from the dynamical center.''  For large $i$, one
expects to see both sides of the jet, and $\Phi_\mathrm{jet}$ is then
about half of the full extent of the radio emission.  However, for
small $i$, $\Phi_\mathrm{jet}$ may not represent the \emph{size} of
the emission region, but its position shift from the actual location
of the dynamical center (the black hole) as the jet becomes more and
more beamed toward the observer.  In this case, the correct comparison
is the core shift between two frequencies predicted by the jet model.
Figure~\ref{fig:jet_model}\textit{b} shows the predicted core shift
between 8.4 and 15~GHz for \NGC{4579}.  Once again, the predicted
shifts are significantly larger than our upper limits for large $i$,
requiring $i \leq 42\degr\ (30\degr)$ for the moderate (\NGC{4258})
model for our 0.3~mas core-shift upper limit.


Table~\ref{tab:gal_size_limits} shows the results for this simple
model analysis for our LLAGN sample at 8.4~GHz.  The
$\Phi_\mathrm{jet}$ values for a jet inclination angle of $60\degr$
are \emph{all} larger than our VLBA size upper-limits.  Our upper
limits to the core size, and our upper limits to any possible core
shift constrain the $\gamma_\mathrm{e} = 300$ jet models to have $i
\lesssim 50\degr$ for all of our galaxies.  Jet models with larger
$\gamma_\mathrm{e}$ require even smaller inclination angles, and jet
models with smaller $\gamma_\mathrm{e}$ are not viable for large
inclination angles.  Essentially, there is no viable range of
parameter space which allows large-inclination angle jets in the
\citetalias{Falcke_B1999} model.  Improvements to the analytic
simplifications at low $\gamma_\mathrm{e}$ values \emph{may} allow the
jets to be smaller than our upper limits, but the required jet power
would then be excessively large.  One possible way to decrease the jet
size while maintaining the basic \citet{Blandford_K1979} conical jet
structure would be to allow the energy density in ions to be much
greater than the energy density in electrons \citep[$\mu_\mathrm{p/e}
\gtrsim 10$, see][]{Falcke_B1995}, something which might occur
naturally if the jet particles originate in an ADAF-like accretion
flow.

\subsection{The Jet Model and Unification}

The result that all six LLAGNs are predicted to have small inclination
angles by the jet model is somewhat puzzling.  The six LLAGNs in this
sample were selected from \citetalias{Ho_U2001} on the basis of their
VLA spectral indices and 5~GHz flux densities.  They are not extreme
in flux density or luminosity, as $\sim 1/4$ of the
\citetalias{Ho_U2001} sample have larger 5~GHz flux densities, and
about $1/3$ of the sample have greater 5~GHz radio powers
\citepalias[see][]{Ulvestad_H2001.0}.  The radio power of the Palomar
Seyfert galaxies is correlated with the [\ion{O}{3}] $\lambda 5007$
luminosity and the FWHM([\ion{N}{2}] $\lambda 6584$) velocity over
several orders of magnitude \citetext{\citetalias{Ulvestad_H2001.0},
  Figures 5 and~6}, suggesting that the variation in the apparent
radio power is caused primarily by sources having different jet
powers, and is not primarily due to an inclination effect.  Therefore,
we do not expect the six LLAGNs in this study to have any preferential
orientation based solely on their apparent radio powers.  For randomly
oriented jet directions, the probability that any single jet has $i
\le 50\degr$ to match the $\gamma_\mathrm{e} = 300$ model is 0.357, or
$2.1\times 10^{-3}$ for all six objects to appear in this orientation.
These probabilities are further reduced if the \NGC{4258} jet model is
used.

Other selection effects, however, may break the assumed random
distribution.  The sources were not selected based on the degree of
compactness on VLA scales from \citetalias{Ho_U2001}, so potentially
high $i$ objects were not excluded.  The sources \emph{were} selected
to have spectral indices $\alpha_{1.4}^{5} \ge -0.35$, so a spectral
index change with luminosity or orientation could bias our object
selection.  The \citetalias{Falcke_B1999} model does predict a
spectral index change with inclination angle --- but the change is
quite small, predicting $\alpha \approx +0.1$ for $i = 25\degr$, and
$\alpha \approx +0.2$ for $50\degr \lesssim i \lesssim 90\degr$.  At
high inclination angles, an absorber located in the plane
perpendicular to the jet axis could conceivably obscure the jet close
to the black hole, where most of the high-frequency radiation is
emitted.  This would leave the low-frequency emission produced further
away from the black hole, resulting in an overall spectrum with
$\alpha < 0$.  However, this absorber would have to be optically thick
up to at least 43~GHz (J.~M.  Anderson, et al., in preparation), and
would require the jet direction to be independent of AGN type as both
Seyfert~1 and Seyfert~2 galaxies have roughly equal numbers of
``steep'' and ``flat'' radio spectra \citepalias{Ulvestad_H2001.0}.

However, for $\alpha > 0$ the Seyfert~1 galaxies in the
\citet{Ho_U2001} do have slightly higher observed radio powers than
the Seyfert~2 galaxies, and hence five Seyfert~1 galaxies and only one
Seyfert~2 galaxy were selected by our flux density limits.  (More
Seyfert~2 galaxies would be included if the limit was decreased to
1~mJy or less.)  The \citetalias{Falcke_B1999} model predicts
increased flux-densities for low-inclination sources, so this bias
would agree with the model if our sources have low jet-inclination
angles.  According to the standard Unification Model for AGNs
\citep{Antonucci1993}, all AGNs are similar intrinsically, but either
show broad emission lines (Type~1) or do not show broad emission lines
(Type~2) depending on the inclination angle of the accretion disk.  An
optically thick torus surrounds the accretion region, and one needs to
look down the accretion axis (assumed to be the same as the jet axis)
in order to see the broad-line region.  According to this model,
Type~1 AGNs should have smaller inclination angles than Type~2 AGNs.
In their analysis of the Palomar sample, \citet*{Ho_FS1997.1} found
that the ratio of Type~2 to Type~1 Seyfert galaxies was 1.4:1, and the
ratio of Type~2 to Type~1 LINER galaxies was 3.3:1.  According to the
unification model, these ratios correspond to half-opening angles for
the torus of 54\degr\ and 40\degr\ for Seyfert and LINER tori,
respectively.  Thus, the unification model predicts small inclination
angles for five of our galaxies, in basic agreement with the jet
model.  Interestingly, \citet{Georgantopoulos_Z2003} suggest that
\NGC{3147}, our only Type~2 AGN, is actually a Type~1 Seyfert galaxy
which lacks a broad-line region.  If true, this would give agreement
between unification and the jet model of the radio emission for all of
our galaxies.

\section{Conclusions\label{sec:conclusion}}

We have performed multifrequency, milliarcsecond-scale radio imaging
of six LLAGNs with flat/inverted spectra.  All six galaxies have
slightly rising spectra throughout the gigahertz range, and are
unresolved on scales of $10^3$--$10^4$ Schwarzschild radii.  The radio
powers are higher than predicted by simple ADAF models, but the radio
powers and spectra are consistent with dominance by compact radio
jets.  However, the unresolved nature of the radio emission implies
that modified ADAF models including nonthermal electrons are still
viable models.  Otherwise, jet models require the jets to be seen
closer to end-on than to side-on in all six galaxies.  This result is
consistent with the fact that at least five of the size galaxies are
Type~1 galaxies, where unified schemes predict that our viewing angle
is within $\sim 45\degr$ of the jet axis.



\acknowledgments

We thank Heino Falcke for several useful discussions on jet models and
LLAGNs.  We would also like to thank Jean Eilek, Sera Markoff, and
Feng Yuan for valuable discussions on the physics of accretion and jet
models.  J.M.A. gratefully acknowledges support from the predoctoral
fellowship program from NRAO.  The work of L.C.H is funded by the
Carnegie Institution of Washington and by NASA grants from the Space
Telescope Science Institute (operated by AURA, Inc., under NASA
contract NAS5-26555).  This research has made use of the SIMBAD
database, operated at CDS, Strasbourg, France, and NASA's Astrophysics
Data System.

\bibliographystyle{apj}

\bibliography{jma_astro}


\clearpage



\plotone{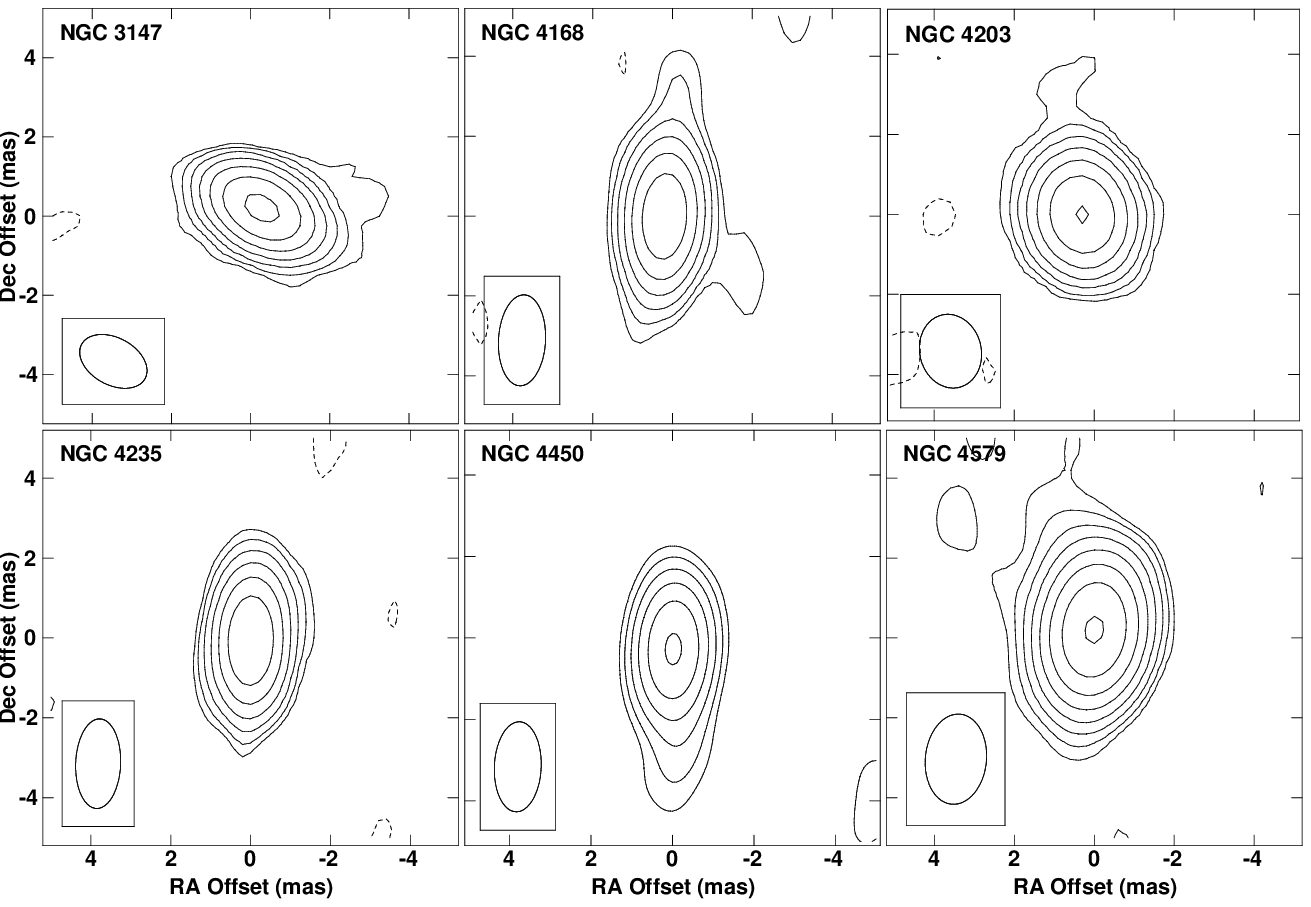} 
\figcaption[f1.eps]{ The
  8.4~GHz VLBA images of the six LLAGNs are shown as contour plots.
  All images are 10.2~mas on a side, with contour intervals increasing
  by factors of 2 starting from two times the rms noise level in each
  image (54, 44, 63, 47, 66, and 59~$\mathrm{\mu Jy~beam^{-1}}$, for
  NGC~3147, NGC~4168, NGC~4203, NGC~4235, NGC~4450, and NGC~4579,
  respectively).  Negative contours are indicated by dashed lines.
  The restoring beam for each image is shown in the lower left-hand
  corner.
\label{fig:8GHz_images}
}

\clearpage

\plottwo{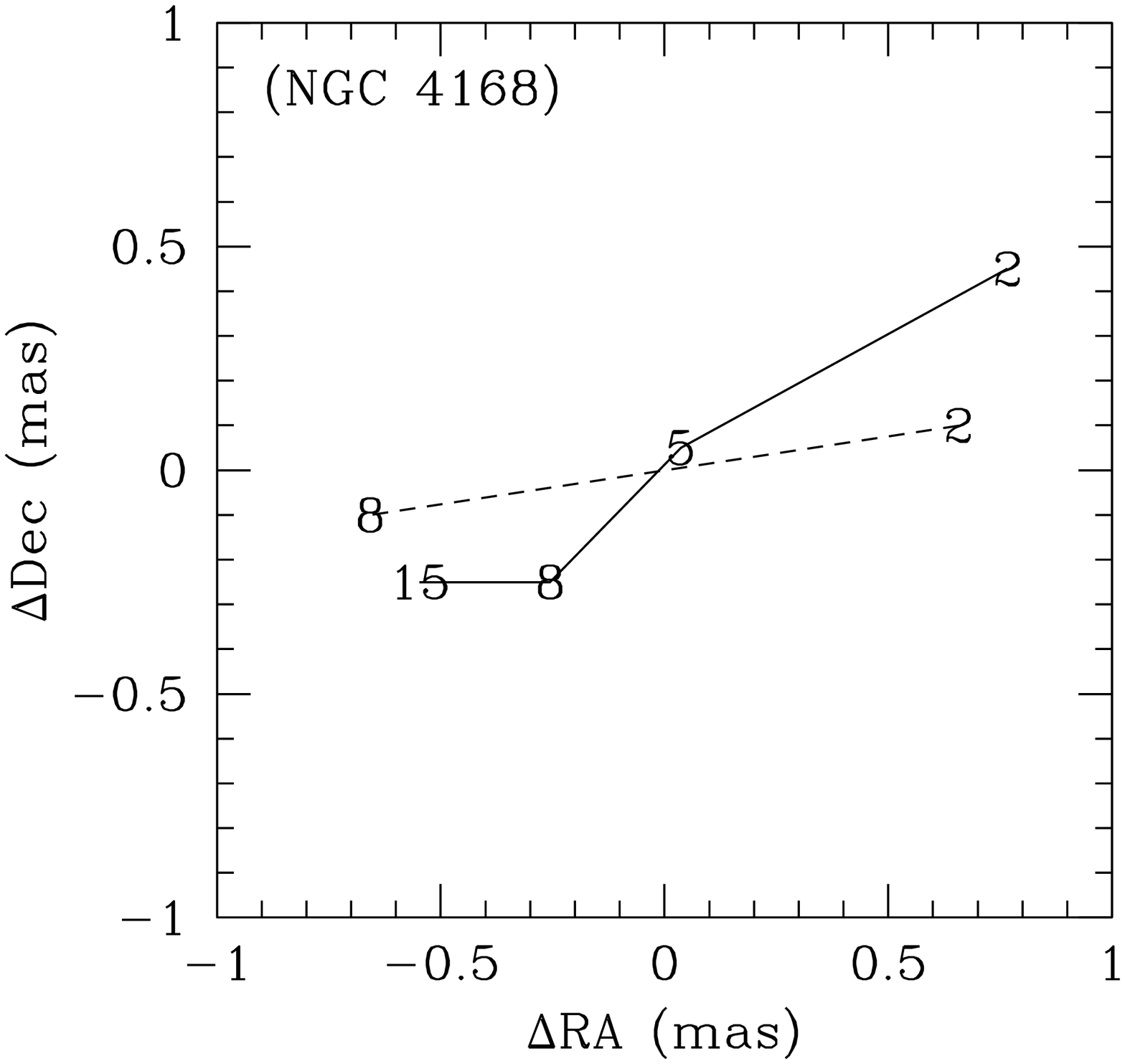}{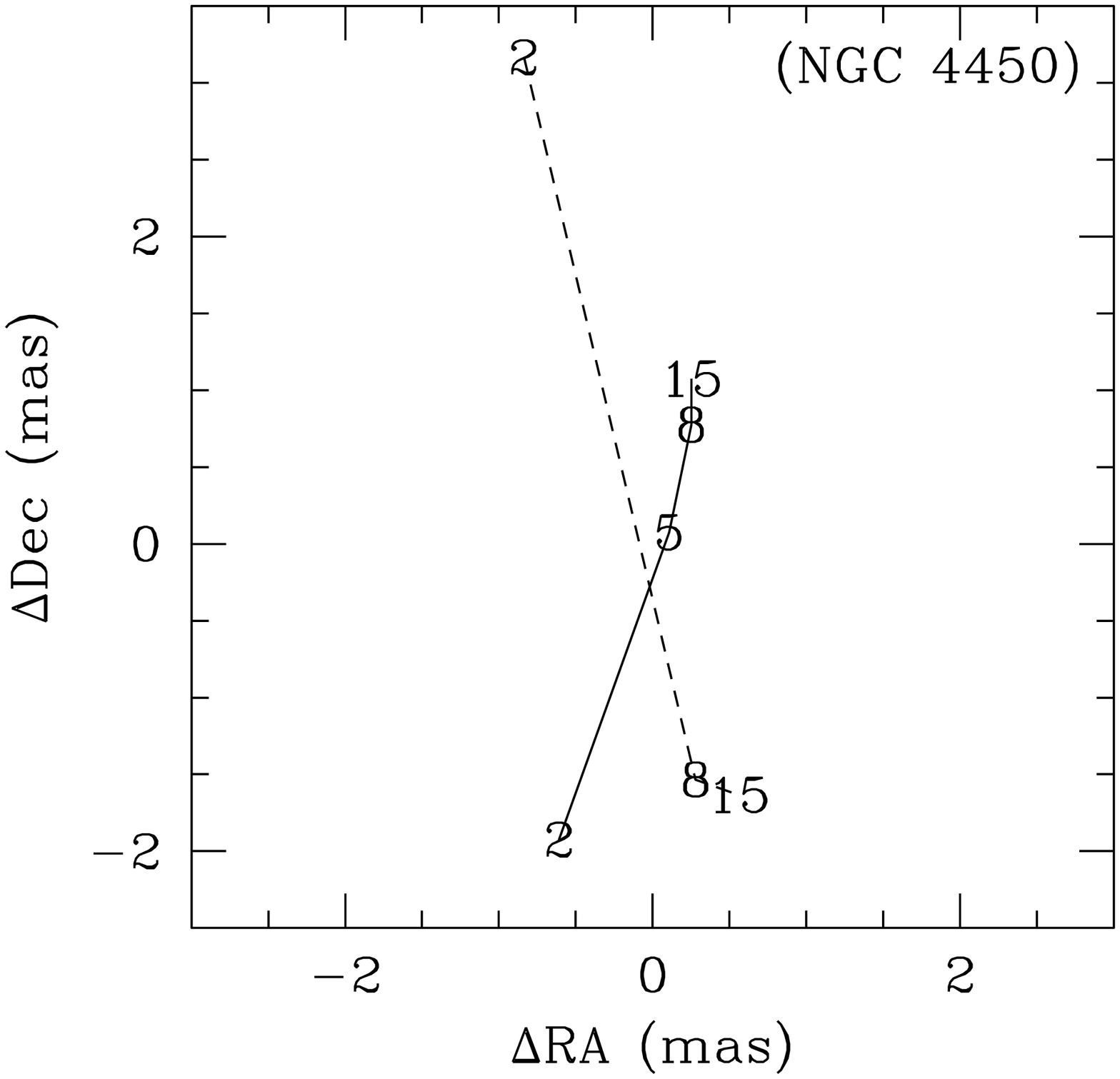}
\figcaption[f2a.eps f2b.eps]{Core
  position is shown as a function of frequency for two target galaxies
  and their check sources.  NGC~{4168} (red) is connected by solid
  lines and its check source (blue) by dashed lines.  NGC~{4450} (red)
  is shown with solid lines and its check source (blue) by dashed
  lines.  Numerals are used to indicate the observing frequency (the
  check sources were not observed at all frequencies).  The measured
  positions have been offset by a constant arbitrary amount for each
  source.  The shift decreases in angular size with increasing
  frequency.  Note that both NGC~{4168} and its check source are
  east-northeast of their phase calibrator, while NGC~{4450} is
  southwest, and its check source northwest of their phase calibrator.
  The angular separations between the phase calibrator and sources are
  about twice as large for NGC~{4450}, probably leading to a larger
  ionospheric position shift.  (See also Tables \ref{tab:obs}
  and~\ref{tab:gal_att}.)
\label{fig:core_shift}
}

\clearpage

\plotone{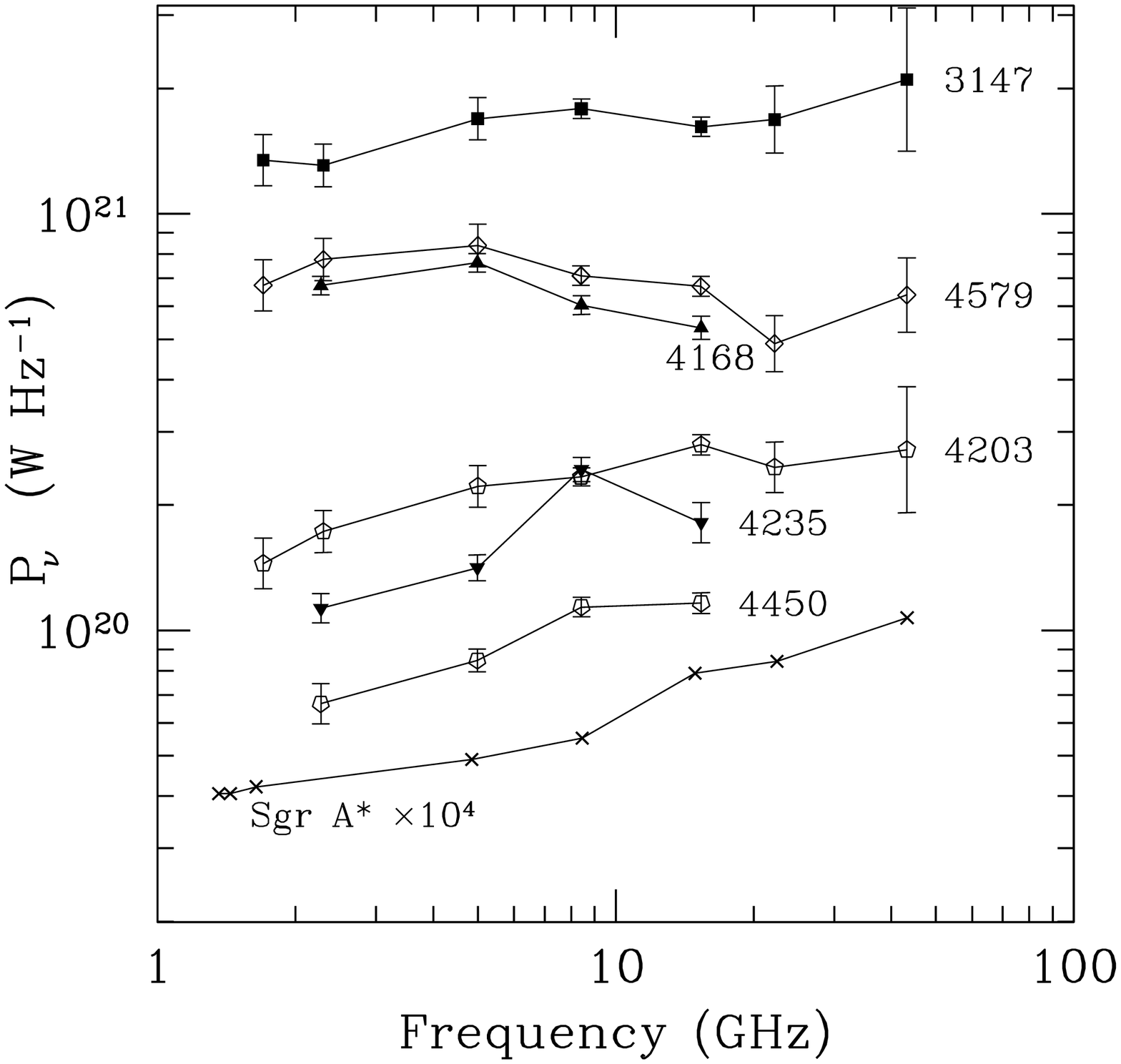} \figcaption[f3.eps]{ The radio powers and
  uncertainties of the target galaxy cores are shown as a function of
  frequency, assuming isotropic radiation.  For comparison, the
  spectrum of {Sgr~A$^\ast$} is shown multiplied by $10^4$
  \citep{Falcke_ea1998}.
\label{fig:power_spectrum}
}

\clearpage

\plotone{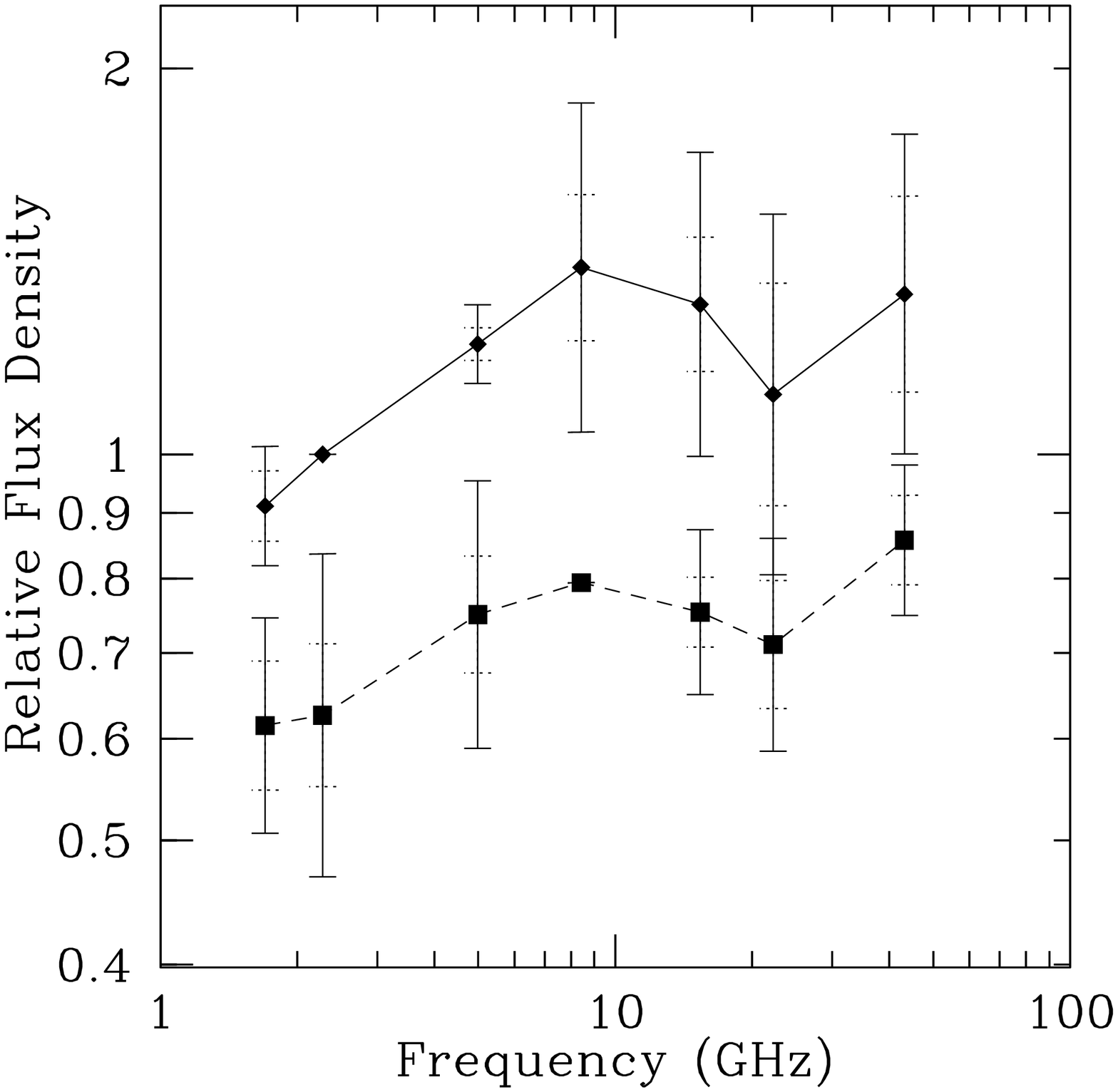} 

\figcaption[f4.eps]{ Mean spectral energy distributions of
  the 6 LLAGNs are shown.  Data points connected by the solid line
  were made after normalizing the spectra to unity at 2.3~GHz.  The
  data points connected by the dashed line were first normalized to
  unity at 8.4~GHz before averaging.  These data points have been
  shifted downward in the figure by 0.1~dex for clarity.  Solid error
  bars show the rms scatter about the mean at each frequency, while
  the dotted error bars closer in represent the statistical
  uncertainty in the mean.
\label{fig:sed}
}

\clearpage

\plotone{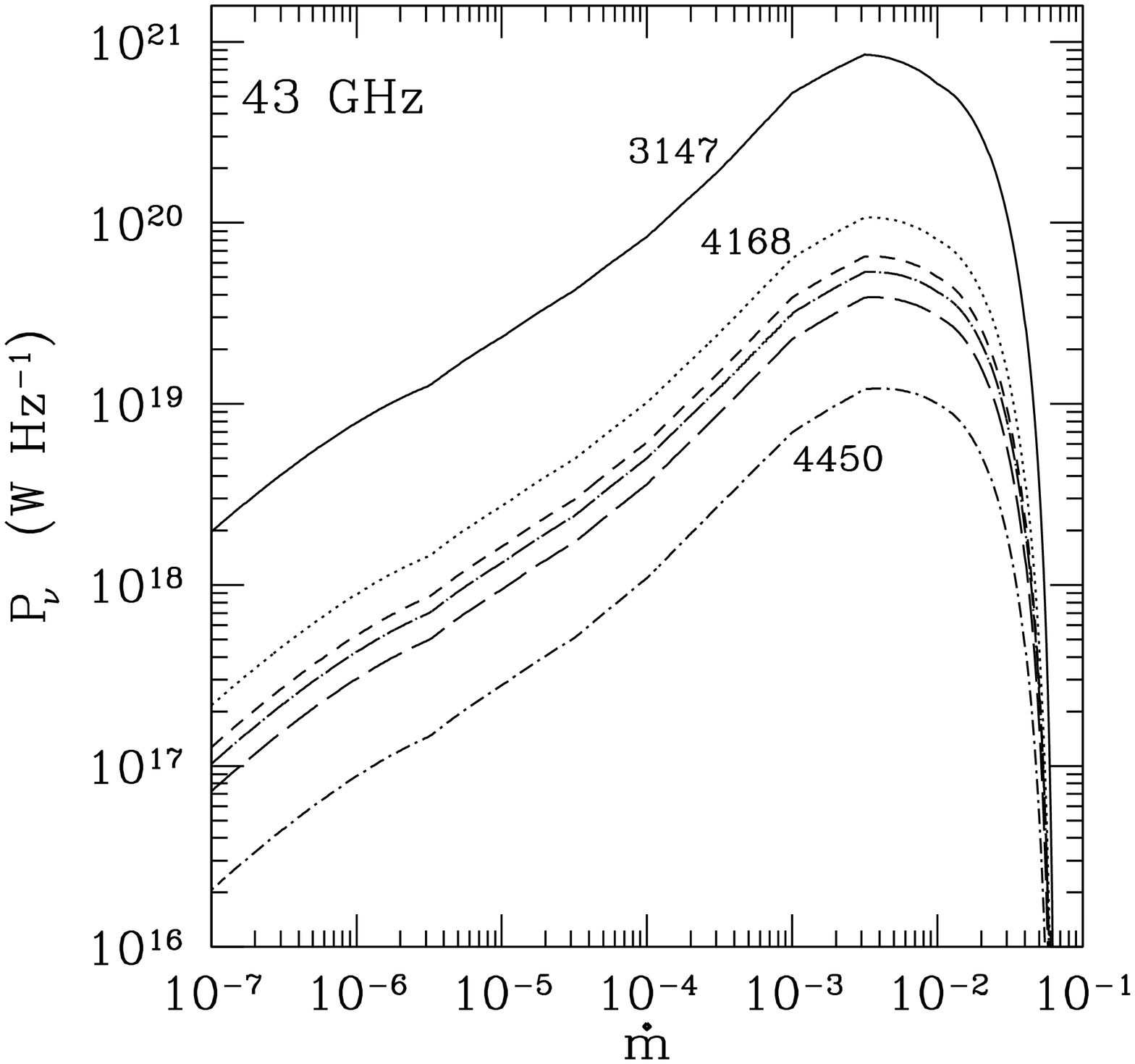}
\figcaption[f5.eps]{The 43~GHz radio power derived
  from \citet{Mahadevan_1997} is shown as a function of the scaled
  accretion rate for the six LLAGNs in this paper.  The emission is
  low for small $\dot{m}$ as there are few particles to radiate.  For
  large $\dot{m}$, the particle density becomes high enough that
  cooling is important, and the electron temperature decreases to a
  point where synchrotron emission is not produced.  NGC~3147 (solid,
  red), with the highest black hole mass of the sample, has the
  correspondingly greatest emission for a given $\dot{m}$.  For
  clarity, NGC~4203 (dashed, blue), NGC~4235 (long dashed, magenta),
  and NGC~4579 (long-dash dot, black) are not labeled on the plot.
\label{fig:ADAF_43}
}

\clearpage

\plotone{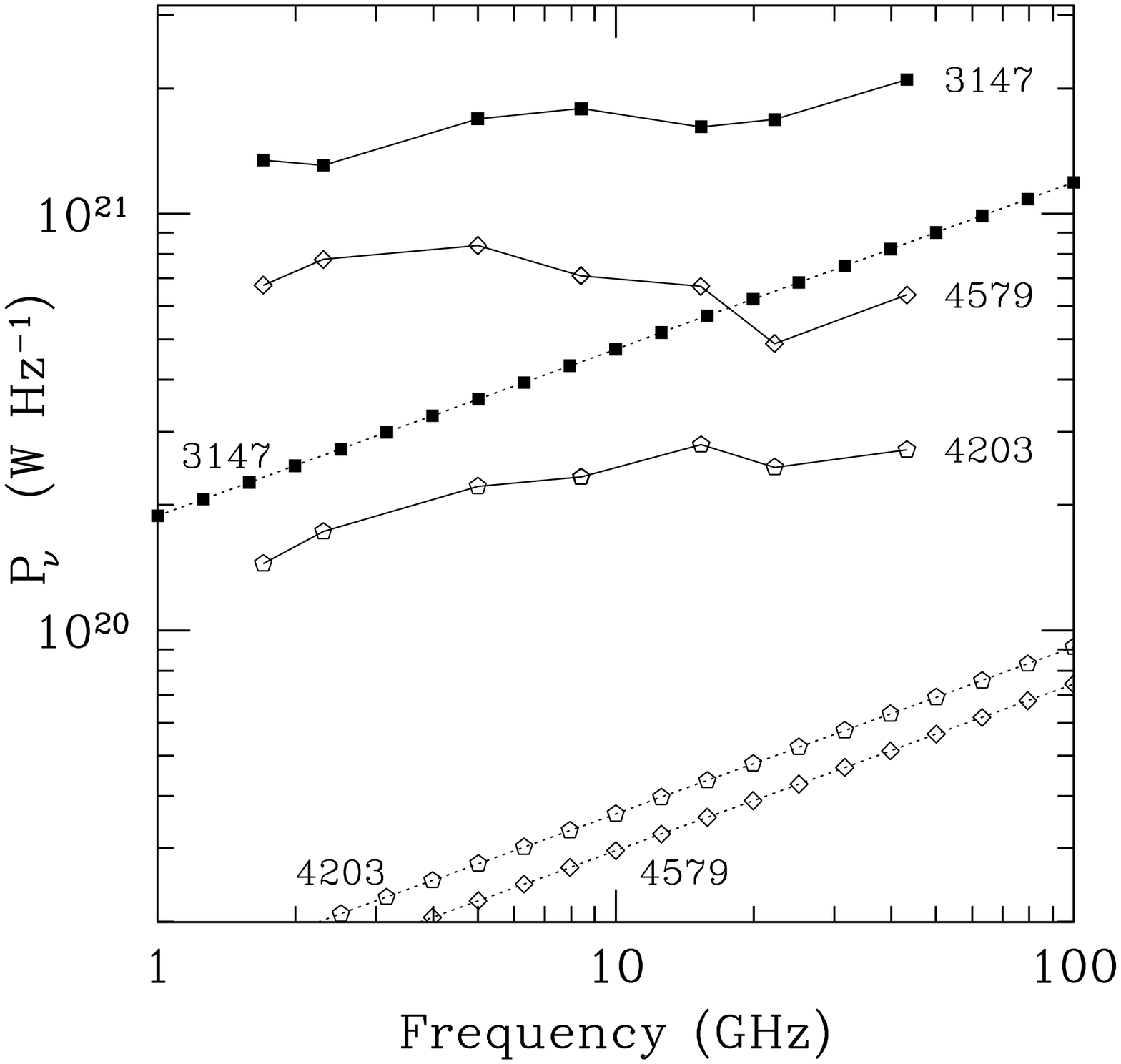}
\figcaption[f6.eps]{
  The maximum radio power from \citetalias{Mahadevan_1997} equation~25 is
  shown as a function of frequency for three LLAGNs in this sample.  The
  spectrum is calculated for $\dot{m} \approx 10^{-2.5}$, which gives the
  maximum radio output in the \citetalias{Mahadevan_1997}
  approximation scheme (see also Figure~\ref{fig:ADAF_43}).  The ADAF
  models are the regularly spaced points with slope $+0.4$ and are
  labeled at the right with their corresponding galaxy.  For
  comparison, the \emph{observed} radio spectra from
  Figure~\ref{fig:power_spectrum} are also shown.  Most of the ADAF
  models fall a factor of $\sim 10$ or more below the observations,
  except for NGC~3147, which is a factor of 2.5 below the 43~GHz
  observation. 
\label{fig:ADAF_spec}
}

\clearpage

\plotone{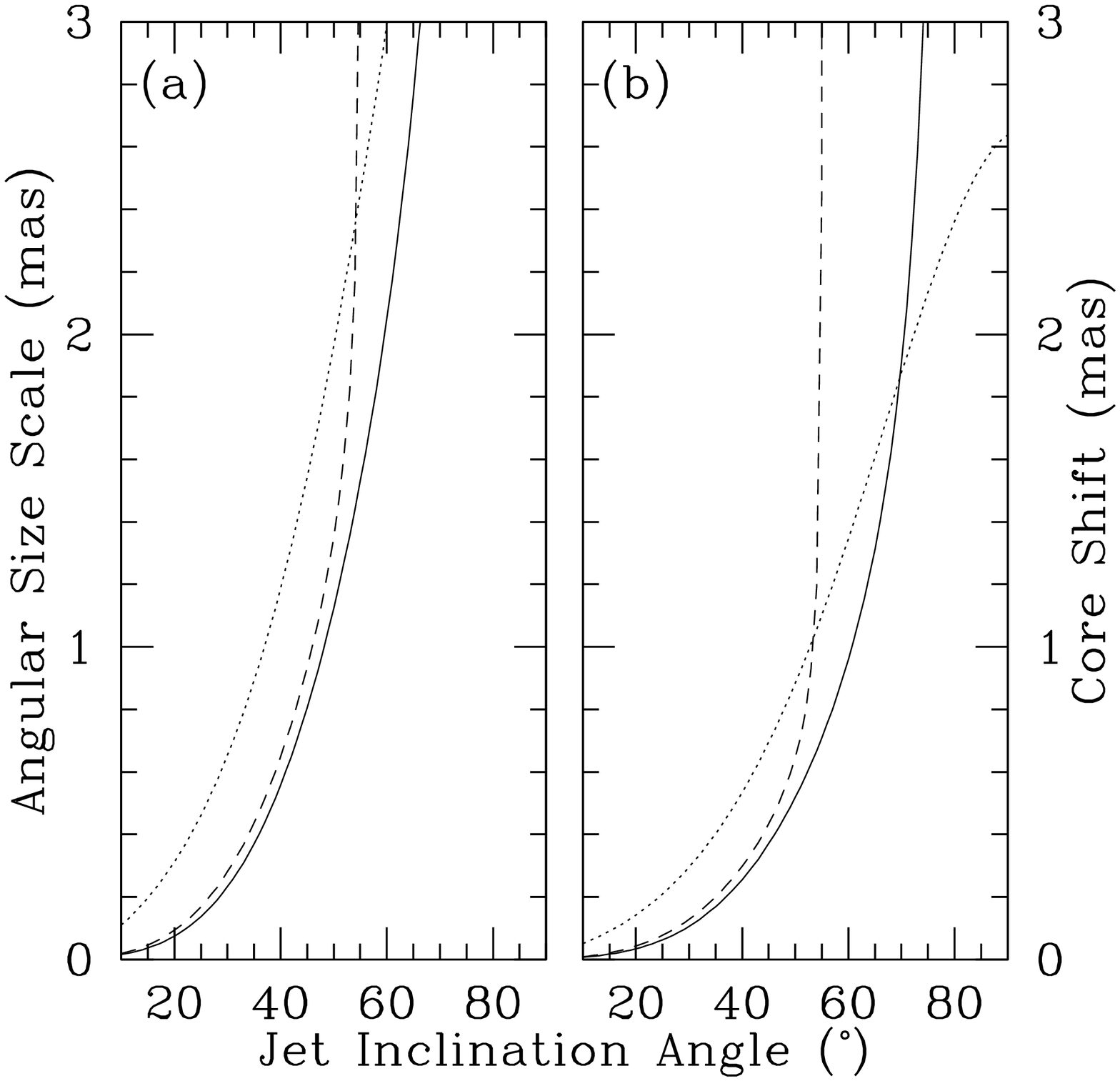} 

\figcaption[f7.eps]{ The angular size scale at
  8.4~GHz ($\Phi_{8.4}$, panel \textit{a}) and core shift ($\Phi_{8.4}
  - \Phi_{15}$, panel \textit{b}) in milliarcseconds are shown as a
  function of jet inclination angle for three jet models of NGC~4579.
  The solid (green) lines correspond to the $\gamma_\mathrm{e} = 300$
  model;  the dotted (red) lines correspond to model parameters
  similar to NGC~4258 ($\gamma_\mathrm{e} = 630$); and the dashed
  (blue) lines correspond to the $\gamma_\mathrm{e} = 250$ model.
\label{fig:jet_model}
}





\clearpage
\begin{deluxetable}{llclcc}
  \tablecaption{VLBA Observation Details\label{tab:obs}}
  \tablewidth{0pt}
  \tablehead{
    \colhead{Galaxy} & \colhead{UT Date} & \colhead{Frequencies}
    & \colhead{Phase} & \colhead{Offset} 
    & \colhead{Check} \\
    \colhead{} & \colhead{} & \colhead{(GHz)}
    & \colhead{Calibrator} & \colhead{(\degr)}
    & \colhead{(\degr)} \\
    \colhead{(1)} & \colhead{(2)} & \colhead{(3)} 
    & \colhead{(4)} & \colhead{(5)} 
    & \colhead{(6)} 
    }
\startdata
\NGC{3147} & 2002 Jun \hfill 21 & 8.42, 15.37, 22.23, 43.22 & 
\objectname[ICRF]{J1048$+$7143} & 2.89 & 3.22 \\
\NGC{4168} & 2002 Jul \hfill 30 & 2.27, \phn 4.99, \phn 8.42, 15.37 & 
\objectname[ICRF]{J1207$+$1211} & 1.60 & 1.80 \\
\NGC{4203} & 2002 May \hfill 11 & 8.42, 15.37, 22.23, 43.22 & 
\objectname[ICRF]{J1220$+$3431} & 1.69 & 0.91 \\
\NGC{4235} & 2002 Aug \hfill 03 & 2.27, \phn 4.99, \phn 8.42, 15.37 & 
\objectname[ICRF]{J1214$+$0829} & 1.40 & 2.76 \\
\NGC{4450} & 2002 Aug \hfill 27 & 2.27, \phn 4.99, \phn 8.42, 15.37 & 
\objectname[ICRF]{J1237$+$1924} & 3.18 & 3.57 \\
\NGC{4579} & 2002 Jun \hfill 02 & 8.42, 15.37, 22.23, 43.22 & 
\objectname[ICRF]{J1230$+$1223} & 1.78 & 5.87 \\
\enddata
\tablecomments{Column~(1): Galaxy name.  Col.~(2): UT observing date.
  Col.~(3): Centers of observed frequencies.  Col.~(4): Calibration
  source for phase referencing.  Col.~(5): Angular separation between
  sources (1) and (4).  Col.~(6): Angular separation between the phase
  calibrator and the check source.}
\end{deluxetable}

\clearpage
\begin{deluxetable}{lccccccccl}
\tabletypesize{\scriptsize}
  \tablecaption{Galaxy Attributes\label{tab:gal_att}}
  \tablewidth{0pt}
  \tablehead{
    \colhead{Galaxy} & \colhead{$\alpha$}
    & \colhead{$\delta$} & \colhead{rms} 
    & \colhead{$D$} & \colhead{Ref.} 
    & \colhead{$\sigma$} & \colhead{Ref.}
    & \colhead{$M_\mathrm{BH}$}  
    & \colhead{Class} \\
    \colhead{} & \colhead{(J2000)} 
    & \colhead{(J2000)} & \colhead{(mas)} 
    & \colhead{$\!\!$(Mpc)$\!\!$} & \colhead{} 
    & \colhead{$\!\!\!\!\!\!(\mathrm{km~s^{-1}})\!\!\!\!\!\!$} & \colhead{}
    & \colhead{(M$_\Sun$)}    
    & \colhead{} \\
    \colhead{(1)} & \colhead{(2)} 
    & \colhead{(3)} & \colhead{(4)} 
    & \colhead{(5)} & \colhead{(6)}
    & \colhead{(7)} & \colhead{(8)} & \colhead{(9)}   
    & \colhead{(10)} 
    }
\startdata
\NGC{3147} & 10 16 53.6503 & $+$73 24 02.696 & 2.3 & 40.9 & 1 & 269 & 2 & $4.4\times 10^8$ & S2 \\
\NGC{4168} & 12 12 17.2685 & $+$13 12 18.701 & 1.4 & 30.9 & 3 & 186 & 4 & $1.0\times 10^8$ & S1.9: \\
\NGC{4203} & 12 15 05.0554 & $+$33 11 50.382 & 1.6 & 15.1 & 3 & 170 & 5 & $7.0\times 10^7$ & L1.9 \\
\NGC{4235} & 12 17 09.8818 & $+$07 11 29.670 & 1.5 & 18.0 & 6 & 155 & 7 & $4.8\times 10^7$ & S1.2 \\
\NGC{4450} & 12 28 29.5908 & $+$17 05 05.972 & 2.1 & 14.1 & 6 & 126 & 8 & $2.1\times 10^7$ & L1.9 \\
\NGC{4579} & 12 37 43.5223 & $+$11 49 05.488 & 2.0 & 19.1 & 6 & 164 & 4 & $6.1\times 10^7$ & S1.9/L1.9 \\
\enddata
\tablecomments{Column~(1) gives the galaxy name, while Columns (2)
  and~(3) give 8.4~GHz VLBA position of the nuclear core (units of
  right ascension are hours, minutes, and seconds, and units of
  declination are degrees, arcminutes, and arcseconds).  Column~(4)
  shows the estimated positional uncertainty in each coordinate,
  including uncertainty in the phase calibrator position, imaging
  noise, and residual phase errors.  Columns~(5) and (6) give the
  adopted distance to the galaxy and a reference for that distance,
  while Columns~(7) and (8) give the adopted velocity dispersion and
  reference.  The resulting mass estimate of the black hole using
  \citet{Tremaine_ea2002} is shown in Column~(9).  Column~(10) gives
  the AGN classification according to \citet{Ho_FS1997.0}, with S
  representing Seyfert galaxies and L representing LINERs.}
\tablerefs{1. \citealt{Tully_1988}, 2. \citealt*{Whitmore_MT1985}, 3.
  \citealt{Tonry_ea2001}, 4.
  \anchor{http://www-obs.univ-lyon1.fr/hypercat/}{HyperLeda} database
  \citep[see][]{Prugniel_ea2001}, 5. \citealt*{Barth_HS2002}, 6.
  \citealt{Solanes_ea2002}, 7.  \citealt{Jimenez-Benito_ea2000}, 8.
  \citealt{McElroy_1995}.}
\end{deluxetable}

\clearpage



\clearpage
\begin{deluxetable}{lrrrrrrr}
  \tablecaption{VLBA Flux Density Measurements\label{tab:flux}}
  \tabletypesize{\small}
  \tablewidth{0pt}
  \tablehead{
    \colhead{Galaxy}
    &\colhead{$S_{1.7}$}
    &\colhead{$S_{2.3}$}
    &\colhead{$S_{5.0}$}
    &\colhead{$S_{8.4}$}
    &\colhead{$S_{15.4}$}
    &\colhead{$S_{22.2}$}
    &\colhead{$S_{43.2}$}
    \\
    &\colhead{(mJy)}
    &\colhead{(mJy)}
    &\colhead{(mJy)}
    &\colhead{(mJy)}
    &\colhead{(mJy)}
    &\colhead{(mJy)}
    &\colhead{(mJy)}
    \\
    \colhead{(1)} 
    &\colhead{(2)}&\colhead{(3)}&\colhead{(4)}
    &\colhead{(5)}&\colhead{(6)}&\colhead{(7)}
    &\colhead{(8)}
    }
\startdata
\NGC{3147} & $  6.7 \pm 1.0 $ & $ 6.5 \pm 0.8 $& $ 8.4 \pm 1.0 $& $ 8.9 \pm 0.5 $& $ 8.1 \pm 0.4 $& $   8.4 \pm    1.7 $& $10.5\phn\pm   5.1 $ \\
\NGC{4168} & \nodata          & $ 5.9 \pm 0.3 $& $ 6.7 \pm 0.4 $& $ 5.3 \pm 0.3 $& $ 4.7 \pm 0.3 $& \nodata             & \nodata              \\
\NGC{4203} & $  5.3 \pm 0.8 $ & $ 6.3 \pm 0.8 $& $ 8.1 \pm 1.0 $& $ 8.5 \pm 0.4 $& $10.2 \pm 0.6 $& $   9.0 \pm    1.3 $& $ 9.9\phn\pm   4.1 $ \\
\NGC{4235} & \nodata          & $ 2.9 \pm 0.3 $& $ 3.7 \pm 0.3 $& $ 6.3 \pm 0.4 $& $ 4.7 \pm 0.6 $& \nodata             & \nodata              \\
\NGC{4450} & \nodata          & $ 2.8 \pm 0.3 $& $ 3.6 \pm 0.2 $& $ 4.8 \pm 0.3 $& $ 4.9 \pm 0.3 $& \nodata             & \nodata              \\
\NGC{4579} & $ 15.4 \pm 2.3 $ & $17.7 \pm 2.2 $& $19.2 \pm 2.4 $& $16.2 \pm 0.9 $& $15.3 \pm 0.9 $& $  11.1 \pm    1.9 $& $14.6   \pm    3.3 $ \\
\enddata
\tablecomments{Column~(1) gives the galaxy name.  Columns (2)--(8)
  give the integrated flux density ($S_\nu$) and uncertainty for
  measurements at 1.7, 2.3, 5.0, 8.4, 15.4, 22.2, and 43.2~GHz. Data
  values for \NGC{3147}, \NGC{4203}, and \NGC{4579} at 1.7, 2.3, and
  5.0~GHz from \citetalias{Ulvestad_H2001.1} have been scaled by
  multiplication to match the more recent 8.4~GHz flux densities.  }
\end{deluxetable}


\clearpage

\begin{deluxetable}{lccccccc}
  \tablecaption{8.4~GHz Size Information\label{tab:gal_size_limits}}
  \tablewidth{0pt}
  \tablehead{
    \colhead{Galaxy} & \colhead{Beam}
    & \colhead{$\theta_{3\sigma}$} & \colhead{$T_\mathrm{b}$} 
    & \colhead{$R$} & \colhead{$\Phi_{60}$}
    & \colhead{$i_{\Phi,\mathrm{max}}$} & \colhead{$i_{\Delta\Phi,\mathrm{max}}$} \\
    \colhead{} & \colhead{(mas)} 
    & \colhead{(mas)} & \colhead{(K)} 
    & \colhead{($R_\mathrm{S}$)} & \colhead{(mas)}
    & \colhead{($\degr$)} & \colhead{($\degr$)} \\
    \colhead{(1)} & \colhead{(2)} 
    & \colhead{(3)} & \colhead{(4)}
    & \colhead{(5)} & \colhead{(6)}
    & \colhead{(7)} & \colhead{(8)}
    }
\startdata
\NGC{3147} & $1.82\times 1.19$ & 0.70 & $> 3.1\times 10^{8}$ &  $<\phn1\,600$ & 1.6 & 47 & 46 \\
\NGC{4168} & $2.28\times 1.18$ & 0.77 & $> 1.5\times 10^{8}$ &  $<\phn6\,000$ & 1.2 & 53 & 49 \\
\NGC{4203} & $1.86\times 1.54$ & 0.59 & $> 4.2\times 10^{8}$ &  $<\phn3\,200$ & 1.5 & 44 & 47 \\
\NGC{4235} & $2.24\times 1.13$ & 0.64 & $> 2.7\times 10^{8}$ &  $<\phn6\,000$ & 1.3 & 48 & 46 \\
\NGC{4450} & $2.21\times 1.15$ & 0.83 & $> 1.2\times 10^{8}$ & $<14\,000$ & 1.2 & 54 & 50 \\
\NGC{4579} & $2.26\times 1.53$ & 0.54 & $> 9.6\times 10^{8}$ &  $<\phn4\,300$ & 2.0 & 40 & 42 \\
\enddata
\tablecomments{Column~(2) gives the size of the restoring beam in
  milliarcseconds.  Column~(3) gives the 3-$\sigma$ upper limit to the
  deconvolved core size along the major axis.  Column~(4) gives the
  corresponding lower limit to the brightness temperature of the
  unresolved core.  Column~(5) gives the corresponding upper limit to
  the \emph{radius} of the unresolved core in Schwarzschild radii.
  Column~(6) gives the characteristic scale size of the 8.4~GHz
  emission for the $\gamma_\mathrm{e} = 300$ jet model from
  \citetalias{Falcke_B1999} for an inclination angle of $60\degr$.
  Column~(7) gives the maximum inclination angle for the
  characteristic size scale to remain less than $\theta_{3\sigma}$ at
  8.4~GHz, and Column~(8) gives the maximum inclination angle which
  results in an 8.4 to 15~GHz core shift of less than 0.3~mas for the
  same model.  For a jet with NGC~4258 model parameters
  \citepalias{Falcke_B1999}, the sizes at $i = 60\degr$ are about 60\%
  \emph{larger}, and the maximum inclination angles allowed by the
  size upper limits are about $13\degr$ \emph{smaller} than the values
  in columns (7) and~(8).}
\end{deluxetable}

\clearpage

\begin{deluxetable}{lcc}

  \tablecaption{VLBA Spectral Indices\label{tab:gal_spec}}
  \tablewidth{0pt}
\centering
  \tablehead{
    \colhead{Galaxy} & \colhead{$\alpha_{2.3}^{5.0}$}
    & \colhead{$\alpha_{8.4}^{15}$} \\
    \colhead{} & \colhead{} 
    & \colhead{} \\
    \colhead{(1)} & \colhead{(2)} 
    & \colhead{(3)} 
    }
\startdata
\NGC{3147} & $+0.33 \pm 0.09$ & $-0.16 \pm 0.13$ \\
\NGC{4168} & $+0.16 \pm 0.09$ & $-0.21 \pm 0.14$ \\
\NGC{4203} & $+0.32 \pm 0.09$ & $+0.30 \pm 0.13$ \\
\NGC{4235} & $+0.29 \pm 0.14$ & $-0.49 \pm 0.23$ \\
\NGC{4450} & $+0.30 \pm 0.17$ & $+0.04 \pm 0.13$ \\
\NGC{4579} & $+0.10 \pm 0.09$ & $-0.10 \pm 0.13$ \\
\enddata
\tablecomments{Radio spectral indices measured with the VLBA are shown
  for individual galaxies.  Column~(2) shows the spectral index from
  2.3 to 5.0~GHz, and Column~(3) shows the spectral index from 8.4 to
  15~GHz.  ($S_\nu \propto \nu^{+\alpha}$.)}

\end{deluxetable}


\end{document}